\documentclass[twocolumn,showpacs,preprintnumbers,amsmath,amssymb,floatfix]{revtex4}
\usepackage{graphicx}

\usepackage[center]{subfigure}
\usepackage{bigints}
\usepackage{cancel}
\usepackage{amsmath}
\usepackage{mathbbol}
\usepackage[center]{subfigure}
\usepackage{tikz}
\usepackage{standalone}
\usepackage{stackengine}

\usepackage{mathtools}

\begin{document}
\author{Sara Najem$^{1}$, Alaa Krayem$^{1}$, Tapio Ala-Nissila$^{2,3}$,  Martin Grant$^{4}$}
%%%
%%
\affiliation{$^1$ Physics Department, American University of Beirut, Beirut 1107 2020, Lebanon}

\affiliation{$^2$ Department of Applied Physics and QTF Center of Excellence, Aalto University, P.O. Box 11000, FI-00076 Aalto, Espoo, Finland}
\affiliation{$^3$ Interdisciplinary Centre for Mathematical Modelling, Department of Mathematical Sciences, Loughborough University, Loughborough, Leicestershire LE11 3TU, United Kingdom}

\affiliation{$^4$ Physics Department, Rutherford Building, 3600 rue University, McGill University, Montr\'eal, Qu\'ebec, Canada H3A 2T8}

\date{May 7, 2020}

\begin{abstract}
We analyze the morphology of the modern urban skyline in terms of its roughness properties. This is facilitated by a database of $10^7$ building heights in cities throughout the Netherlands which allows us to compute the asymptotic height difference correlation function in each city. We find that in cities for which the height correlations display power-law scaling as a function of distance between the buildings, the corresponding roughness exponents are commensurate to the Edwards-Wilkinson and Kardar-Parisi-Zhang equations for kinetic roughening. Based on analogy to discrete deposition models we argue that these two limiting classes emerge because of possible height restriction rules for buildings in some cities.
%We also recover the distribution of the species which represent buildings from different construction periods along with their allometric scaling laws. 
 \end{abstract}
%\title{The Urban Skyline as a Moving Interface: An Indication of KPZ-EW Cross-over}%City's Buildings as Competing Species and 
\title{Kinetic roughening of the urban skyline}%City's
\maketitle
%\section{Introduction}

There are three major strands of urban morphological analysis all of which give insight into the dynamical process of urban change  \cite{:wp}. These respectively study the micro-morphology, the relation between morphological periods and the typological process, and finally the relation between decision making and urban form. 
The first follows the intra-city spatial change, which has been shown to be clustered over time and diffuses spatially  \cite{:wp}.
The second strand lays out the typological process of change in buildings types where the latter are viewed as resulting from a process of learning from adaptations of the previous building types. It also follows how morphological characteristics are superseded by those of the next. Last, the interplay between decisions has been shown to lead to either intended or unintended fringe belts around preserved historical zones as well as to delineating different morphological periods  \cite{:di,Whitehand:1977en}. 

The amount of change in urban form has been linked to the neighborhood effect, which is itself dependent on to the dwelling density; in high density environments the changes tend to be imitative \cite{:wp}.
Moreover, the typological process was investigated by H. Bernoulli who came up with the notion of property cycles, which follows the gradual filling of a lot of land by buildings which in due course are replaced with newer ones when their life cycle comes to an end \cite{:un}. 
The cycle is divided into Boom,  Slump and Recovery phases \cite{:wp}. The high density housing prevails in the booming phase while the interplay between low land values and the geographical constraints leads to fringe belts, which include vegetation areas, landmark, buildings of architectural importance. The belt thus forms a boundary zone between historically and morphologically distinct housing areas \cite{:di,Whitehand:1977en}. 

The city concept has been studied through the lens of statistical physics as the dynamical processes at play in urban allometry, mobility, urban form and social segregation, to list a few, and has parallels in the study of magnetic materials, phase transitions, the Ising model and many others \cite{bettencourt2013origins,krug1997origins,castellano2009statistical,simini2012universal,barthelemy2019statistical,barthelemy2016structure,batty2008scaling,batty2008size,louf2013modeling,bettencourt2007growth,atis2015experimental}. The modern urban skyline, being an important city metric in the assessment of the city's solar energy, visual complexity,  {and urban climatology, particularly the effect of its roughness on scalar transfer coefficient \cite{heath2000tall,calcabrini2019simplified,chung2015wind,hagishima2009aerodynamic,zaki2011aerodynamic,ikegaya2012geometric}}, has been as well followed empirically; however no dynamical description has been provided to explain its evolution  \cite{schlapfer2015urban}. 

\begin{figure}[!htp]%  
  %\includestandalone{latexEx2}
    \includegraphics[scale=0.4]{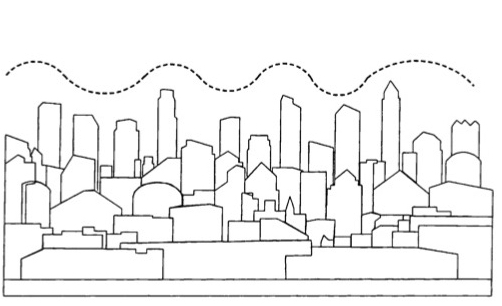} 
  \caption{A schematic of the urban skyline modeled as a moving interface. The figure is adapted from Ref. \cite{nasar2002skyline}.}
  {\label{fig:skylinescheme}}
\end{figure}

Under the effect of the dynamical processes described above, such as property cycles and the spatial diffusion of morphological changes, the buildings' heights are constantly varying with alternating growth and decay linked to construction and destruction. Thus the local height function $h(\vec{r},t)$ of the city at position $\vec{r} = (x,y)$ at time $t$, can be thought of as a dynamic, spatio-temporally evolving stochastic quantity describing growth phenomena (cf. Fig. \ref{fig:skylinescheme}). This is reminiscent of interface dynamics problems commonly studied in contexts of film growth, flame front propagation, particle deposition, turbulent liquid crystals, growth of bacterial colonies, and directed polymers in random media just to list a few \cite{halpin1995kinetic,corwin2012kardar,kim1989growth,almeida2014universal,atis2015experimental,halpin20122,de2017memory}.  In such systems, macroscopic observables, such as height fluctuations and relevant correlation functions often exhibit power-law dependence on system size $L$ and on time due to kinetic roughening of the height correlations \cite{halpin2014universal,ala1993scaling,meerson2016large}.

If the individual heights of the buildings evolved stochastically and independently without any bounds, the interface dynamics would be that of a simple random deposition (RD) model, for which the surface width $w^2(L,t) = \int d\vec{r} [h(\vec{r},t) - \bar{h}(t)]^2/L^{d_{s}}$ simply grows as $w(t) \propto t^{1/2}$ in time. Here $\bar{h}(t)$ is the spatially averaged instantaneous height over a system of linear size $L$, and $d_{\rm{s}}=2$ is the spatial dimension of the front. For cases where there are nontrivial correlations between the heights, the Family-Viscek scaling ansatz is often obeyed as
\begin{equation} \label{firstscaling}
w(L,t) \propto L^{\alpha} f(\frac{t}{L^z}),
\end{equation}
where the scaling function
\begin{equation} \label{scalingfunction}
f(x) \propto \begin{cases} x^{\beta}, & \mbox{for } x \ll 1;
\\ {\rm const.}, & \mbox{for } x \gg 1. \end{cases}
\end{equation}
The quantities $\beta$, $\alpha$ and $z=\alpha/\beta$ define the growth, roughness and dynamical scaling exponents, respectively. These can also be determined from the height difference pair correlation
function \cite{ala1993scaling} 
 
{which is commonly used to measure the standard deviations in the underlying height probability distribution function as
\begin{equation} \label{scalingofG}
G({r},t) = \left\langle[h({\vec r} + \delta {\vec r},t) - h({\vec r},t)]^2 \right\rangle_r ,
\end{equation}
}
where $\langle \cdot \rangle_r$ denotes (isotropic) averaging over space. The asymptotic limit of the 
 {
isotropic}
$G(r,t)$ is given by
\begin{equation} \label{correlationscaling}
G_{\infty}(r,t) \propto \begin{cases} r^{2\alpha}, & \mbox{if } r_c \ll t_{\infty}^{1/z}; \\ t_{\infty}^{2\beta}, & \mbox{if } r_c \gg t_{\infty}^{1/z}, \end{cases}
\end{equation}
where $t_{\infty}$ and $r_c$  are respectively the saturation time and saturation distance of the function; that is the time and distance at which the function plateaus 
 
{which measures the extent of correlations between the heights.}
%In order to calculate $G_{\infty}(r)$, the pairwise distance $r_{ij}$ between all locations $i$ and $j$ is needed in order to compute the height difference, as in Equation \ref{scalingofG}, for all the points falling within $r + dr$.% where $dr$ was chosen to be a belt of of $50m$.
More complicated anomalous and multiscaling behavior has also been reported in some models
\cite{asikainen2002interface}.

Perhaps the best known examples of models of kinetic roughening following Family-Viscek scaling
are the nonlinear Kardar-Parizi-Zhang (KPZ) equation and its linear Edwards-Wilkinson (EW) counterpart \cite{kardar1986dynamic,edwards1982surface}. 
The KPZ equation includes a term playing the role of surface tension in the Hamiltonian picture, a non-linear driving force that accounts for slope-dependent growth velocity, and a stochastic term, respectively, given by
 \begin{equation} \label{kpz}
\frac{\partial h }{\partial t} = \nu \nabla^2 h  + \frac{\lambda}{2} (\nabla h)^2 +  
\eta(\vec{r},t), % \frac{\lambda}{2} (\nabla h)^2
\end{equation}
where $\eta$ is uncorrelated Gaussian noise whose Fourier transformation satisfies $\langle \hat{\eta}(\vec{k},t) \hat{\eta}(\vec{k}',t') \rangle = D (2 \pi)^{d_{\rm s}} 
\delta^{d_{\rm s}} (\vec{k} + \vec{k}') \delta(t-t')$, and $\vec{k}$ is the momentum. It reduces to the exactly solvable linear EW equation when $\lambda = 0$, for which the growth exponents are $\alpha = \beta = 0$ (logarithmic), and $z = 2$ for $d_{\rm s}=2$ 
\cite{majaniemi1996kinetic}.
For the KPZ case, the exponents are known exactly only in $d_{\rm s} =1$, and in $d_{\rm s} =2$ they have been numerically estimated to be $\alpha \approx 0.39$, $\beta \approx 0.24$, and $z = 2 - \alpha \approx 1.61$ 
\cite{ala1994scaling}.

In this work we utilize a large database of $\approx 10,000,000$ building heights in cities throughout the Netherlands \cite{bag3d} to study the roughness of the height function for each city. Such height functions form compact two dimensional surfaces that are naturally single-valued due to the lack of voids and overhangs. The building heights are expected to be somehow correlated due to structural engineering constraints and possible city specific zoning regulations that may restrict both the absolute heights and height differences between nearby buildings. Thus, we expect to find well-defined scaling exponents at least in some cases, and their values should reflect the influence of the different scenarios.

%Further, making assumptions about $\beta$ we could recover cities' ages and compare them with their actual. %Our motivation is grounded on the  %However, unlike the KPZ equation the heights are stabilized and subsequently we propose a variation to it which offers an explanation to the evolution of the urban skyline. 
%The exponent characterizing the growth phase through the scaling of $w(t) \propto t^{\beta}$ and $w(t) \propto  L^z$ reveal the underlying ``mircoscopic" processes at play. We propose the following modifications to the KPZ equation which account for a ``gravity term" and a dynamic effect given respectively by $g$ and $v$ and show that we can recover the exponents of the Manhattan data through this governing equation: 

The dataset that we use includes city names, their buildings' footprints, construction years, and heights at given coordinates.
The \textit{bag3d geopackage} dataset was split according the attribute NAMES$\_2$, corresponding to the city's name, which allowed us to generate a separate \textit{geopackage} for 491 cities. The buildings' centroids were retrieved for them in order to compute the pairwise distance $r$ between buildings which is necessary to compute $G(r,t)$. While it is easy to consider the spatial correlations, 
the data represent the cities' current configurations and age distributions, and therefore
the time variable (growth dynamics) is not well defined. Thus, we have to assume that the correlation function has already saturated in $2020$ and we then compute $G_{\infty}(r)$ to extract the roughness exponent $\alpha$.
%For example, buildings' centroids falling within the Amsterdam boundaries are tagged with the city label in order to carry out the analysis within its bounds and repeat the process to other cities.

\begin{figure}[!htp]%
  %\includestandalone{latexEx2}
    \includegraphics[scale=0.13]{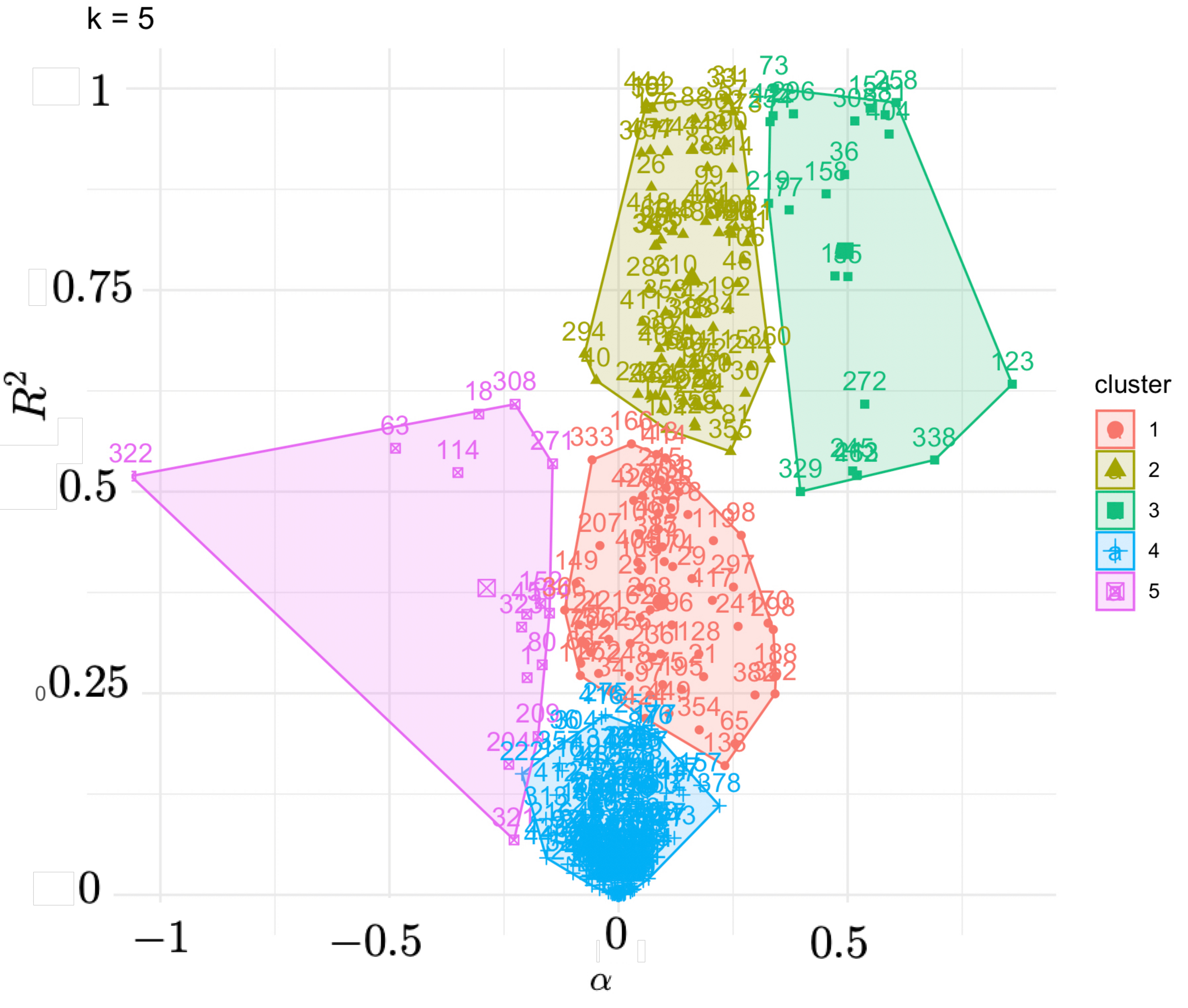}
  \caption{Clustering of the cities according to their corresponding values of $\alpha$ and $R^2$ from a linear fit of the $G_{\infty}(r)$ versus $r$ on a logarithmic scale.}
  {\label{clusters}}
\end{figure}

\begin{figure}[!htp]
  \subfigure[]  {\label{fig:GofKPZ}}%{8cm}
    \includegraphics[width=7.2cm]{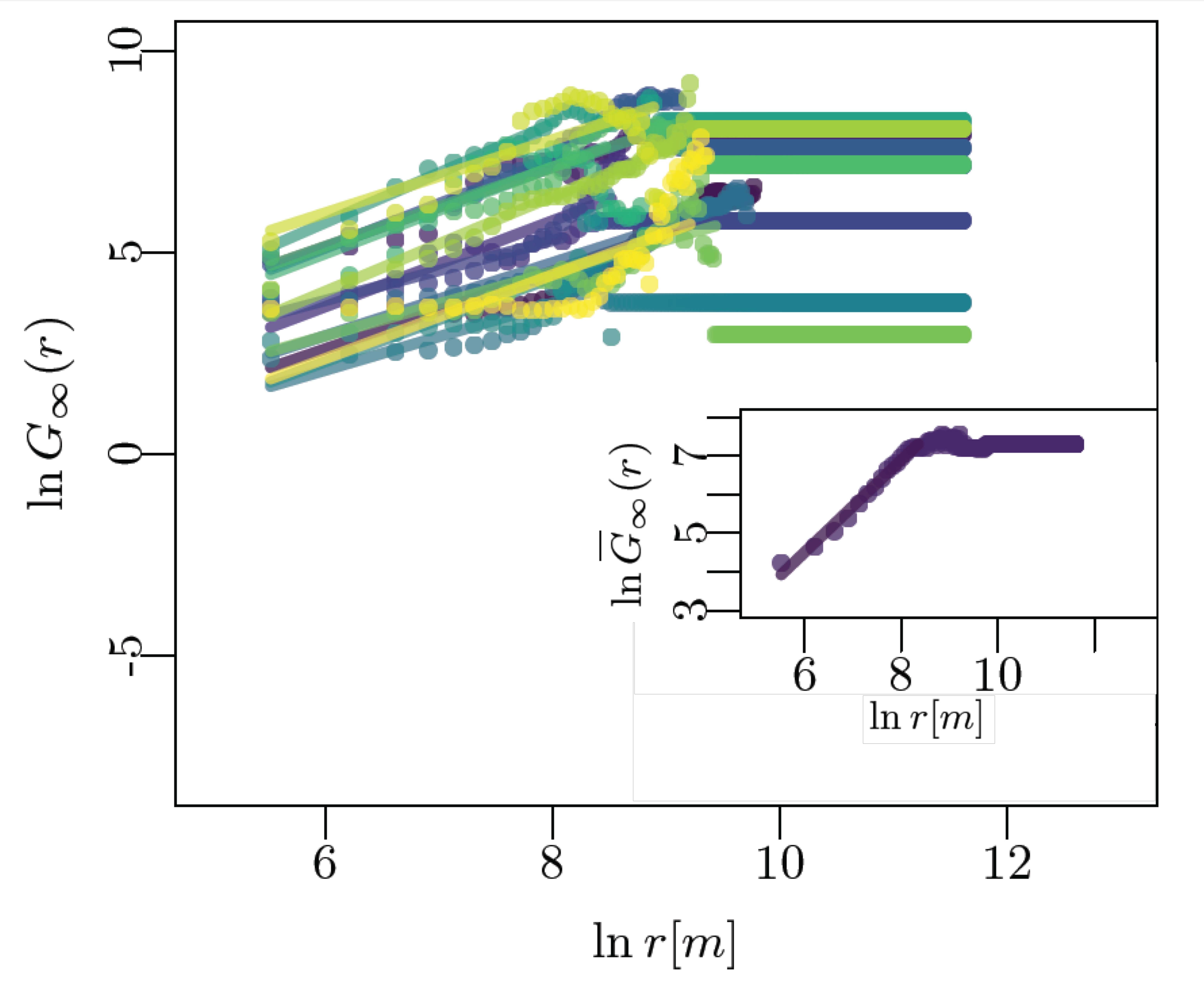}%GofRKPZ.pdf
    \\
            \subfigure[]{\label{fig:GofEW}}%{8cm}
    \includegraphics[width=7.2cm]{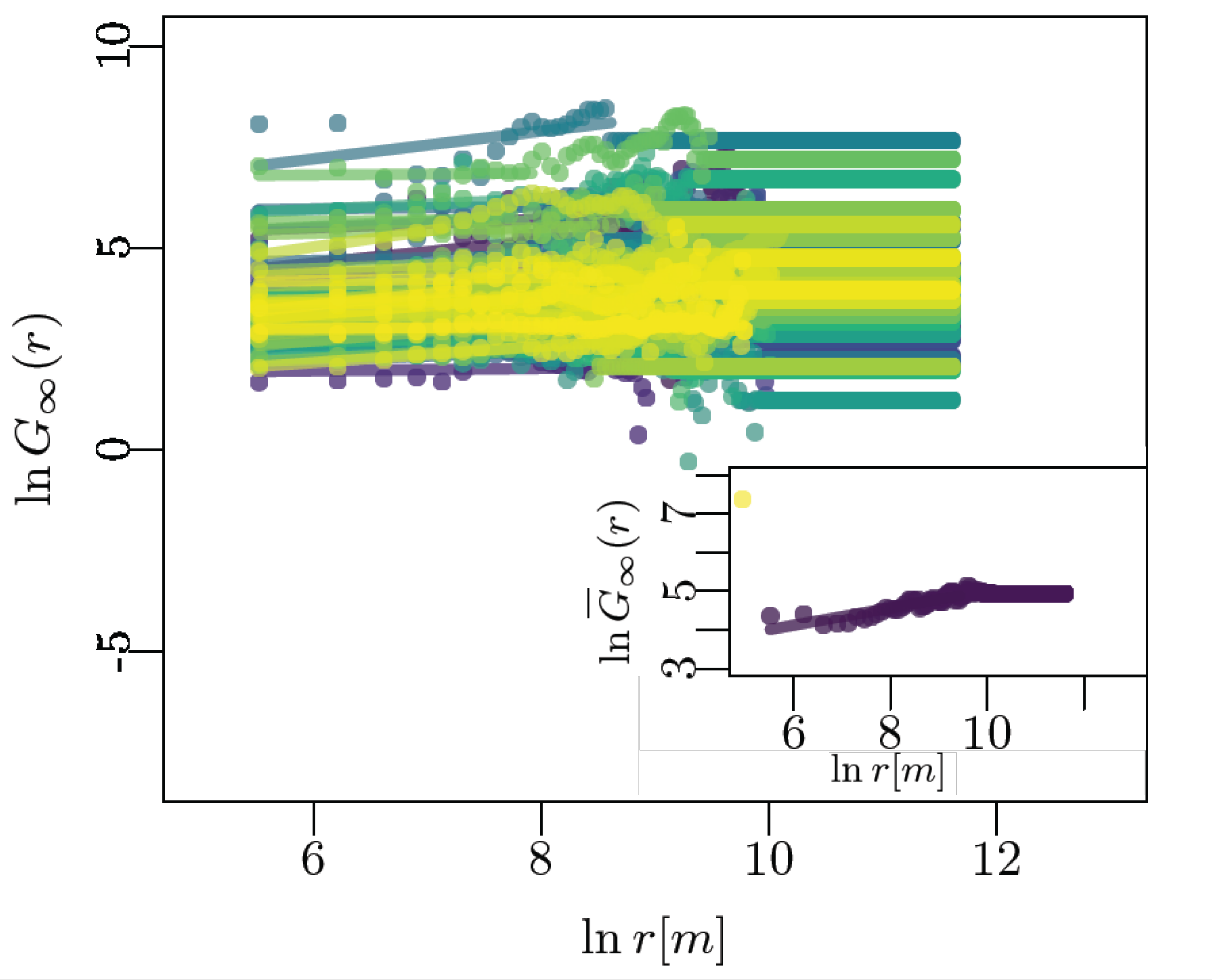}%GofEW2
     \caption{Saturated spatial correlation functions $G_{\infty}(r)$ for the two clusters with high values of $R$-squared. In (a) we show the functions for cluster no. 3 of 15 cities for which $\alpha = 0.47 \pm 0.09$, while in (b) the functions are shown for cluster no. 2 of 117 cities with $\alpha = 0.05 \pm 0.05$. The insets show data for the averaged correlation function of the cities analyzed in each cluster. The correlation function data have been shifted vertically for clarity of presentation. See text for details.}
         \label{fig:discrete}
\end{figure}

Our code was parallelized to run on GPUs and $G_{\infty}(r)$ was computed for each city. We immediately found that, although there is some degree of crossover apparent, the cities can be reasonably clustered according to their values of $\alpha$ and the goodness of the linear fit, measured by the $R^2$ value of fitting to $\log{G_{\infty}(r)}$ versus $\log{r}$. The clustering is shown in Fig. \ref{clusters}. We note that clusters $1,4,5$ have a low $R^2$ and thus we neglect them from further analysis. We focus on the cities belonging to clusters $2$ and $3$, which comprise 117 and 15 cities respectively. The correlation functions $G_{\infty}$ for these cities are shown in Figs. \ref{fig:GofEW} and \ref{fig:GofKPZ} respectively. 

The inset shows the average $G_{\infty}$ for both clusters. The data indicate a relatively well-defined power law dependence on the distance, with saturation occurring at about $8$ km. In each case shown here we included only cities that had at least 10 data points in the power law regime. 
The corresponding average roughness exponents in the two clusters are given by  $\bar{\alpha} = 0.05 \pm 0.05$ (${R}^2 = 0.76 \pm 0.13$), and $\bar{\alpha} = 0.47 \pm 0.09$ (${R}^2 = 0.79 \pm 0.18$).
This indicates that the cities with high confidence in the fitting to the correlation function vary between two classes (there is considerable variation at the highest confidence levels close to $R^2 = 1$), where within the error bars the scaling exponent $\alpha$ is consistent with the EW equation for cluster no. 2 ($\alpha = 0$), and the KPZ equation for cluster no. 3 ($\alpha \approx 0.4$), although for the latter case the average exponent is somewhat larger than the actual KPZ value.  {It is worth mentioning that the 15 cities that fall in the KPZ class here account for 168,321 buildings and extend over an area of $367$ km$^2$}.
%
% using Eq. \ref{correlationscaling}. 
%
%

% \begin{figure}[!htp]%
%  \begin{subfigure} {\label{fig:GofKPZ}}%{8cm}
%    \centering\includegraphics[width=7.2cm]{GofRKPZ.pdf}
%   %\caption{  }
%
%  \end{subfigure}
%    \begin{subfigure}{\label{fig:GofEW}}%{8cm}
%    \centering\includegraphics[width=7cm]{GofEW2}
%% \caption{}
%  
%  \end{subfigure}
%\caption{In the Fig. \ref{fig:GofKPZ} $2\alpha= 0.95   \pm 0.19 $, or equivalently  $\alpha = 0.47 \pm 0.09$, while in Fig. \ref{fig:GofEW} $2\alpha =   0.11 \pm  0.96 $; that is $\alpha = 0.05 \pm 0.05$.  117 cities fell in the EW class while 14 in the KPZ. The rest of the cities belonged to the clusters with $\bar{\alpha} \approx 0$ but their respective $R-$squared was less than 0.5.  }
%\end{figure}
%

\begin{figure}[!htp]%
  %\includestandalone{latexEx2}
    \includegraphics[scale=1]{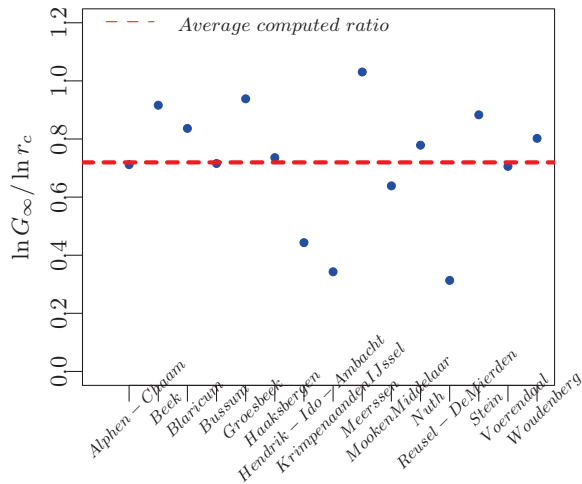}
  \caption{The ratio $\ln{G_{\infty}}/\ln{r_c} = 2\beta z\ln{A} \approx {\rm const.}$ 
from the correlation functions for cities in cluster no. 3. The red line is the average value of $0.76$, which gives $A \approx 2.2$ using $\alpha=0.47$.}
  {\label{IndirectBetaCities}}
\end{figure}

However, the value of the exponent alone is not sufficient to determine whether or not these equations are relevant for the city skyline roughness. For the KPZ equation there is a universal amplitude ratio that can be used to identify the universality class \cite{ala1993scaling}. To this end
we note that for the cities in cluster no. 3 the computation of $G_{\infty}(r,t)$ for $r_c \gg t_{\infty}^{1/z}$ requires knowing $t_{\infty}$, which is available from our data through $r_c$ from the relation $t_{\infty} \propto r_c^{z}$.  The correlation function in time scales as $G_{\infty}(t) = B t_{\infty}^{2\beta}$ which then allows determination of the non-universal amplitude $A$ associated with  $G_{\infty}(r) = A r^{2\alpha}$ \cite{ala1993scaling}. Thus, in the saturated regime $\ln{G_{\infty}}/\ln{r_c} = 2\beta z\ln{A}$ should be a constant. In Fig. \ref{IndirectBetaCities} we show the ratio $\ln{G_{\infty}}/\ln{r_c}$ from the individual cities' correlation functions for cluster no. 3. Although there is some scatter in the data, we find that the ratio is roughly constant and its average value is $0.76$, which gives $A \approx 2.24$ in qualitative agreement with values found for discrete growth models in the KPZ class 
 {
\cite{halpin20122,ala1993scaling}.
}
%
%Now we can estimate the saturation time $t_{\infty}$ up to a proportionality constant $c$, and thus estimate the city age $t_{age} = t_{current} - t_{\infty}$, where  $t_{current}= 2020$ is the year up to which the cities' data are available. They are summarized in Table \ref{cityage}.
%Alphen-Chaam, Bussum, and Meerssen were first mentioned in 711, 1470, and 870  respectively. Other cities, such as Blaricum and Valkenberg, date back to the 10th and 12th centuries respectively. No information could be found regarding the ages of the remaining cities in Table  \ref{cityage}.

\begin{figure}[!htp]%
  %\includestandalone{latexEx2}
    \includegraphics[scale=0.13]{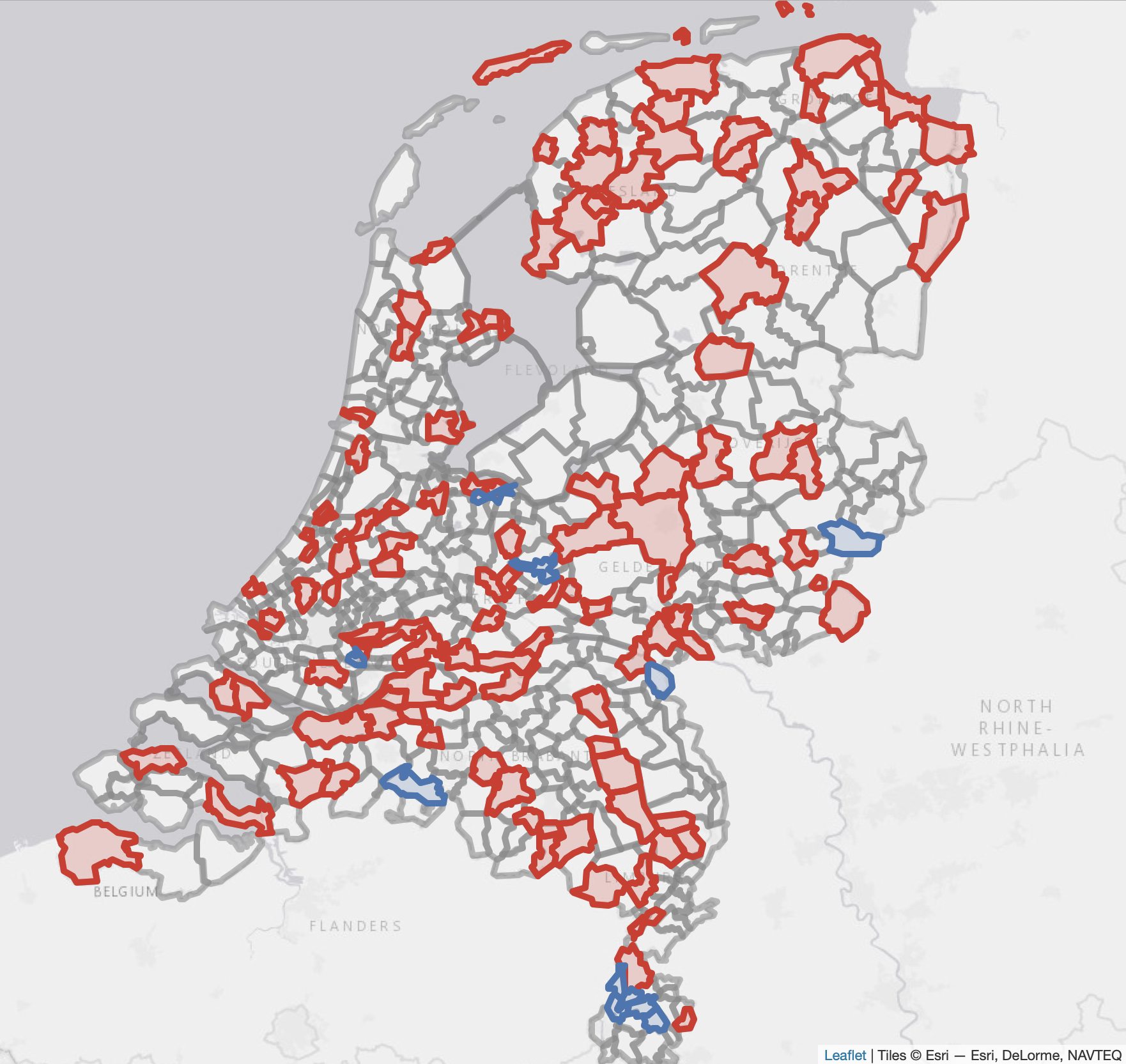}
  \caption{The cities from cluster no. 3 whose skyline roughness is commensurate with the KPZ equation are shown in blue, while those from cluster no. 2 that display EW type of behavior are shown in red. For the rest of the cities in the database the $R^2$ values are two low to reliably indicate power law behavior of the correlation function.}
  {\label{citiesonmap}}
\end{figure}

 In Fig. \ref{citiesonmap} we show the spatial distribution of the cities in the Netherlands from clusters 2 and 3 whose skyline roughening is commensurate with in the two possible EW (red) and KPZ (blue) universality classes.
%
% \begin{figure}[!htp]%
%  %\includestandalone{latexEx2}
%    \includegraphics[scale=0.08]{citiesonmap.jpg}
%  \caption{     The cities which fall in the KPZ universality class are shown in color on top of the Netherlands' boundary shown in blue.   }
%  {\label{citiesonmap}}
%\end{figure}
% \begin{figure}[!htp]%
%  %\includestandalone{latexEx2}
%    \includegraphics[scale=0.08]{citiesonmapEW.jpg}
%  \caption{     The cities which fall in the KPZ universality class are shown in color on top of the Netherlands' boundary shown in blue.   }
%  {\label{citiesonmap}}
%\end{figure}
%
%
% \begin{figure}[!htp]%
%  %\includestandalone{latexEx2}
%    \includegraphics[scale=0.08]{Amsterdam.jpg}
%  \caption{     $ 1.071429    \pm 0.1124616 $	    }
%  {\label{amsterdam}}
%\end{figure}
% \begin{figure}[!htp]%
% \vspace{-1.5cm}
%  %\includestandalone{latexEx2}
%    \includegraphics[scale=1]{Fig1.eps} 
%\vspace{-1.8cm}
%  \caption{ The base 10 logarithm of Eq. $2$, shown in blue,  is fitted to the city's potential and the total road network length, with R-Square $= 0.9437$ and a p-value $= 2.814\times10^{-6}$.}
%  {\label{fig:cityscalingL}}
%\end{figure}
% \begin{figure}[!htp]%
%  %\includestandalone{latexEx2}
%    \includegraphics[scale=1]{SpatialCorrelation} 
%  \caption{  2.3998       0.2585 }
%  {\label{fig:cityscalingL}}
%\end{figure}
% \begin{figure}[!htp]%
%  %\includestandalone{latexEx2}
%    \includegraphics[scale=1]{SpatialCorrelationSuperPc} 
%  \caption{     3.9647       0.1153  }
%  {\label{fig:cityscalingL}}
%\end{figure}
%
%
% \begin{figure}[!htp]%
%  %\includestandalone{latexEx2}
%    \includegraphics[scale=1]{TimeCorrelationSuperPC}
%  \caption{     4.76908      0.04386    }
%  {\label{fig:cityscalingL}}
%\end{figure}
% Please add the following required packages to your document preamble:
% \usepackage[normalem]{ulem}
% \useunder{\uline}{\ul}{}
%
To understand why kinetic roughening equations of the EW or KPZ type could be relevant to the roughness in urban city skylines, it is instructive to look at relevant discrete deposition models that are in these two classes. Perhaps the simplest such models for the present case are the RD model with surface relaxation (RDSR) and restricted solid-on-solid (RSOS) models in the EW and KPZ classes, respectively \cite{katzav2004connection,corwin2012kardar,park1995exact,ala1992driven,buceta2014revisiting}.
Figure 6 shows schematically how interface roughness evolves in these two models. In the RDSR model the particles randomly deposited can relax to their nearest neighbor sites in the lattice if these sites are lower in height, while in the RSOS class of models there's a strict height difference restriction between nearest neighbors.
%
% \begin{figure}[!htp]%
% \begin{subfigure} {\label{rd}}%{8cm}
%    \centering\includegraphics[width=5cm]{rsosdeposition.eps}   %\caption{  }
%    % \vspace{-1.cm}
%  \end{subfigure}
% 
%    \begin{subfigure} {\label{rsos}}%{8cm}
%    \centering\includegraphics[width=5cm]{rsosRD.eps}
%% \caption{}
%  \end{subfigure}
%  \caption{  The scheme of RD is shown in \ref{rd}, while that of the restricted solid on solid deposition RSOS is shown is \ref{rsos} \cite{kim1989growth}.}
%   \label{fig:discrete}
%\end{figure}
%
  \begin{figure}[!htp]
         \centering
 \subfigure[] {\label{rd}}%{8cm}
    \includegraphics[width=3cm]{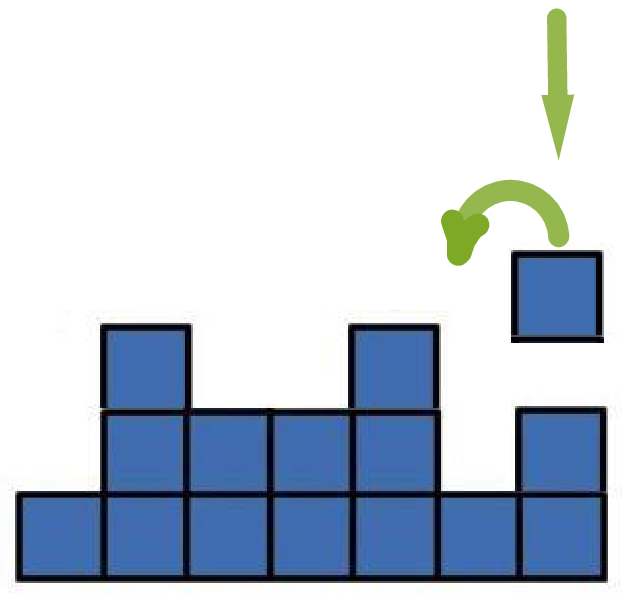}   
       \subfigure[]{\label{rsos}}%{8cm}
    \includegraphics[width=3cm]{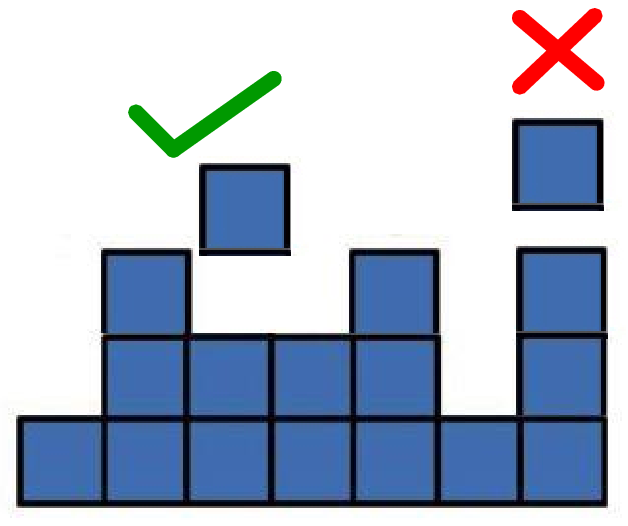}
     \caption{A schematic of the deposition rules in (a) the RDSR and (b) in the restricted solid-on-solid model with nearest neighbor height differences $\vert \Delta h \vert \le 2$ \cite{kim1989growth}. The former is in the EW class while the RSOS models have slope-dependent growth velocity leading to a finite value of $\lambda$ in the KPZ equation.}
         \label{fig:discrete}
\end{figure} 

Based on the above we speculate that the difference in the behavior of the asymptotic correlation function $G_{\infty}(r)$ is related to urban planning constraints as we have also explored the dependence of $\alpha$ on a handful of additional geospatial factors, including the area of the city, its perimeter, and the density of the built environment, all of which proved to have no correlation to the observed scaling. The simplest explanation why EW or KPZ type of roughness may evolve in city skylines is based on building height restrictions: without any explicit restrictions the skyline tends to form a somewhat smoothed EW type of an interface due to natural construction and urban zoning related factors, while strict height restrictions lead to correlations akin to those in the RSOS model.  However plausible, we have not been able to identify any explicit building code height restrictions between these populations in the data.

To summarize, in this work we have considered the morphology of the urban skyline in terms of kinetic roughening of growing fronts. A huge database of about $10,000,000$ building heights in cities throughout the whole of Netherlands has allowed us to analyze the morphology of the city skylines in terms of their roughness exponent $\alpha$. Interestingly enough, for the cases where there is a relatively well-defined power law behavior of the height difference correlation function, we find that the exponents observed fall into and in between two categories which seem commensurate with the EW and KPZ universality classes. A qualitative explanation why this is the case is based on natural smoothing of the skyline for the EW class, and explicit height restriction rules set in some cities which may lead in KPZ type of correlations between the heights.
% \begin{figure}[!htp]%
%  %\includestandalone{latexEx2}
%    \includegraphics[scale=1]{BetaSuperPCKPZCities}
%  \caption{     $t \propto L^z$ ,$z=  0.22  \pm  0.04  $}
%  {\label{fig:cityscalingL}}
%\end{figure}

\section*{Acknowledgment}
T.A-N. is in part supported by the Academy of Finland through its QTF Centre of Excellence program (project no. 312298) and the PolyDyna project (no. 307806). M.G. is supported by the Natural Sciences and Engineering Research
Council of Canada and by {\it le Fonds de recherche du Qu\'ebec 
Nature et technologies}.
%\section*{References}
%\newpage
%\newpage
\bibliography{skyline}
\end{document}